# Time-resolved fuel injector flow characterisation based on 3D laser Doppler vibrometry

**Cyril Crua and Morgan R Heikal**

Centre for Automotive Engineering, University of Brighton, UK

**Abstract**

Hydrodynamic turbulence and cavitation are known to play a significant role in high-pressure atomizers, but the small geometries and extreme operating conditions hinder the understanding of the flow's characteristics. Diesel internal flow experiments are generally conducted using x-ray techniques or on transparent, and often enlarged, nozzles with different orifice geometries and surface roughness to those found in production injectors. In order to enable investigations of the fuel flow inside unmodified injectors, we have developed a new experimental approach to measure time-resolved vibration spectra of diesel nozzles using a 3D laser vibrometer. The technique we propose is based on the triangulation of the vibrometer and fuel pressure transducer signals, and enables the quantitative characterisation of quasi-cyclic internal flows without requiring modifications to the injector, the working fluid, or limiting the fuel injection pressure. The vibrometer, which uses the Doppler effect to measure the velocity of a vibrating object, was used to scan injector nozzle tips during the injection event. The data were processed using a discrete Fourier transform to provide time-resolved spectra for valve-closed-orifice, minisac and microsac nozzle geometries, and injection pressures ranging from 60 to 160 MPa, hence offering unprecedented insight into cyclic cavitation and internal mechanical dynamic processes. A peak was consistently found in the spectrograms between 6 and 7.5 kHz for all nozzles and injection pressures. Further evidence of a similar spectral peak was obtained from the fuel pressure transducer and a needle lift sensor mounted into the injector body. Evidence of propagation of the nozzle oscillations to the liquid sprays was obtained by recording high-speed videos of the near-nozzle diesel jet, and computing the fast Fourier transform for a number of pixel locations at the interface of the jets. This 6–7.5 kHz frequency peak is proposed to be the natural frequency for the injector's main internal fuel line. Other spectral peaks were found between 35 and 45 kHz for certain nozzle geometries, suggesting that these particular frequencies may be linked to nozzle dependent cavitation phenomena.

(Some figures may appear in colour only in the online journal)

## 1. Introduction

The importance of fuel atomisation is well recognized and has been extensively studied experimentally and theoretically [1–11]. It is generally accepted that a key factor in simulating engine combustion processes is the accurate modelling of the interaction between injector nozzle, fuel sprays and in-cylinder flows. The complexity of the processes involved in the injection of diesel fuels is such that many facets involved are still not well understood. Obtaining direct information on the injector's internal fuel flow is particularly challenging due to the small scale of the mechanical design and high flow velocities and pressures. Researchers often resort to applying non-intrusive optical diagnostic techniques onto specially modified [12] or enlarged nozzles, in order to study internal flows and cavitation processes [13, 14]. This method is experimentally challenging and only provides qualitative data, thus hindering the validation of numerical models. Due to the materials and manufacturing techniques used to produce transparent nozzles, their orifices necessarily have a different surface roughness and different inlet geometries compared to production nozzles, both of which are known to affect internal flow turbulence and cavitation.

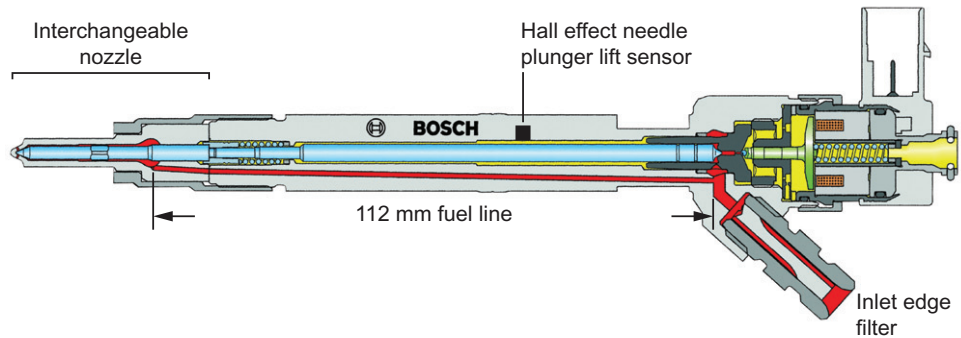

**Figure 1.** Cross-section of a second generation Bosch common rail injector (Picture: Bosch).

The transient nature of the cavitation phenomenon has been emphasized in experimental studies of cavitating flows in cylindrical orifices [15–18] and Venturi nozzle [19, 20], and linked to the process of growth and collapse of vapour bubbles. The intrinsic instability of partially cavitating flows originates from the instability of re-entrant jet motion in the flow separation region [21, 22]. Due to this transient and periodic nature, cavitating flows generate noise and vibration.

There is also some experimental and numerical evidence that high frequency pressure fluctuations exist inside modern diesel injectors, and that they influence the shape and stability of the sprays [23–25]. These fluctuations are usually recorded using pressure sensors fitted to the common-rail or close to the injector. Although these measurements provide insightful information, acoustic attenuation may conceal the link between upstream pressure fluctuations and nozzle cavitation.

The authors previously reported on the observation of high frequency oscillations in needle lift traces, and near-nozzle diesel jet [26]. Oscillations of the injector needle in the axial direction were observed between 6 and 7 kHz, and were found to correlate with the transversal oscillations of the liquid fuel jets. These oscillations were suggested to contribute to the formation of fuel droplet clusters that were seen detaching from the spray. The droplet grouping patterns were predicted by CFD and shown to be in agreement with high-speed visualisations [27]. In order to further investigate the nature and source of diesel nozzle vibrations, an experiment was devised to measure the vibration of the injector tip throughout the injection event, for a range of injection pressures and nozzle geometries. The authors believe that such non-intrusive, quantitative, and time-resolved measurements of high-frequency diesel nozzle vibrations are novel, and could lead to new insight into the internal dynamics of the injector's mechanical components and fuel flow inside unmodified nozzles.

## 2. Experimental setups and data acquisition

### 2.1. Fuel injection equipment

Injections were performed in a static enclosure at atmospheric conditions to minimise the complexity of the experiment and to provide a large optical access. The fuel injection equipment (FIE) consisted of a second generation Bosch common-rail system capable of achieving injection pressures of 160 MPa. It comprised a low pressure pre-supply pump, a fuel filter, a high-pressure pump connected to the common rail, and an injector with interchangeable nozzles. The high-pressure pump was a three-plunger radial-piston pump, operated at a constant speed of 2820 rpm. The common rail was fitted with a pressure sensor and regulated by a pressure regulator valve. The high-pressure fuel line was instrumented with a Kistler 4067 piezoresistive high-pressure transducer. A current clamp was used to record the instantaneous current drawn by the injector. The high-pressure pump and injector were controlled by a purpose-built controller and software that allowed independent control of the injection pressure and timing. The controller allowed triggering the injection event and acquisition chain independently.

The fuel was a low sulphur reference fuel representative of automotive diesel with a density of 830 kg m$^{-3}$ and sulphur content of 0.02% by mass. The fuel injector studied with the vibrometer was a production Bosch CRI 2.2 servo-actuated solenoid common rail injector (figure 1). This type of injector is non-ballistic, hence the injector's needle plunger may reach hydraulic stop if the injection duration is sufficiently long, as can be the case for medium and high engine load [28]. The injector nozzles were interchangeable and therefore allowed testing of different nozzle types and nozzle orifice diameters. The same injector body, including the needle plunger, was used for all vibrometer measurements and only the nozzle tip and needle were exchanged. This approach ensured that spectral measurements could not be affected by mechanical or hydraulic differences between different injectors. A second identical injector was specially instrumented with a Hall effect needle plunger lift sensor to obtain time-resolved position of the needle during the injection. Needle lift traces were recorded for all nozzle types and for a range of injection pressures. A typical needle lift trace is shown in figure 2. In order to allow a comparative investigation of the effect of nozzle type and orifice diameter, the nozzle library included eight single-orifice nozzles with valve-closed-orifice (VCO), minisac and micro-sac geometries (figure 3). The nozzles were custom-manufactured by Bosch with the cylindrical holes spark-eroded and then 10% honed, similarly to standard production nozzles. The orifice was angled at 65° relative to the injector axis. The nozzle diameters and hole length to diameter ratios are listed in table 1. All eight nozzles were tested at six injection pressures ranging from 60 to 160 MPa in 20 MPa increments, giving a total of 48 individual test conditions.

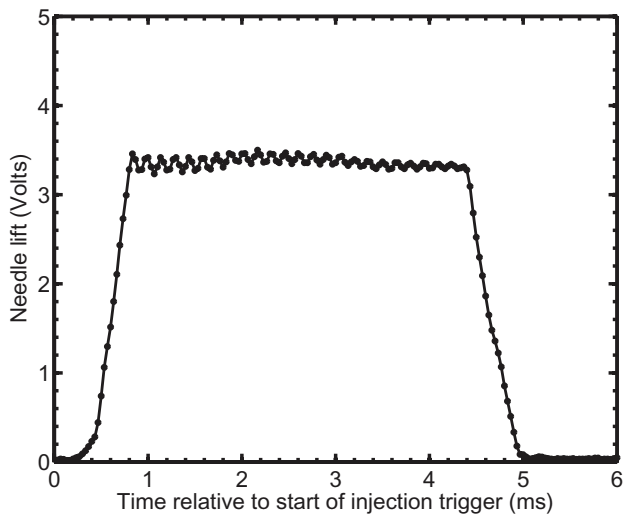

**Figure 2.** Injector needle plunger lift signal for a VCO 200 $\mu$m single-hole nozzle, with an injection pressure of 160 MPa.

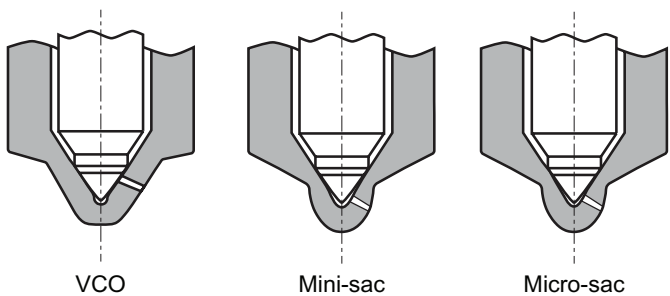

**Figure 3.** Schematics of valve-closed-orifice (VCO), mini-sac and micro-sac nozzle geometries. The holes of a VCO nozzle are closed by the needle, whereas the holes of a sac-type nozzle lead to a single sac which is itself closed by the needle. The micro-sac nozzle has a reduced sac volume compared to the mini-sac design.

### 2.2. Laser Doppler vibrometer

The experimental set-up for the time-resolved 3D laser Doppler vibrometry is shown in figure 4. The laser Doppler vibrometer [29] was a Polytec PSV-400-3D Scanning Vibrometer, composed of three separate scanning vibrometer sensor heads. The three heads each contained a helium-neon (He-Ne) laser that was focused on the same location, over a sample area in the region of 1 mm$^2$. Each head functioned as an independent laser interferometer that derived the instantaneous velocity of the nozzle tip by measuring the Doppler frequency shift of the back-scattered He-Ne laser beam. The vibration components could then be calculated in 3D from the angles of incidence of the lasers. A spatial calibration was required to establish the physical positions of the three LDV heads with respect to the injector nozzle. The relatively small size of the nozzle tip (7 mm diameter) meant that the surface area was not sufficient to perform the spatial calibration with satisfactory accuracy.

**Table 1.** Orifice length to diameter ratios for the injector's single-hole nozzles.

| Orifice diameters | 100 $\mu$m | 160 $\mu$m | 200 $\mu$m |
| --- | --- | --- | --- |
| VCO nozzles | 10 | - | 5 |
| Mini-sac nozzles | 9 | 5.625 | 4.5 |
| Micro-sac nozzles | 10 | 6.25 | 5 |

Hence the 3D spatial calibration was performed on a purpose-built aluminium target plate that could be slipped around the injector nozzle for calibration. The target plate presented a 2D surface of 48 × 30 mm that was designed to be flush with the tip of the injector. Calibration in the third dimension was facilitated by a 20 mm disk surface recessed by 2.5 mm relative to the front plane. Once spatially calibrated, the geometry data of the injector nozzle tip was measured in 3D using the manufacturer's geometry scan unit (PSV-A-420). The calibration of the spatial coordinate system of the LDV was conducted using the manufacturer's software package, and repeated prior to each test session. The validity of the calibration and geometry data was inspected by scanning the nozzle tip surface and checking for signs of misalignment between the three laser beams.

In order to find the optimum spatial location on the nozzle tip for the LDV measurements, nine individual positions were probed and measurements were repeated 30 times to obtain statistically representative spectra. The spectra obtained for the nine spatial locations had similar peak frequencies, but the signal-to-noise ratio was the highest on the tip of the injector's nozzle. This indicates that the nozzle vibrated homogeneously and that the collection efficiency of the LDV was the highest when probing the nozzle's flat tip, which offered a more favourable scattering angle.

Experiments were then conducted in sessions of 3 to 4 h, and a spatial calibration of the LDV system was performed before every session. For each nozzle geometry listed in table 1, LDV measurements were performed at the nozzle tip, and repeated 5 times for each injection pressure ranging from 60 to 160 MPa in 20 MPa increments. The vibrometer was also set to record the instantaneous fuel line pressure and current drawn by the injector. Unless otherwise mentioned, all spectrograms shown were computed using the average of 5 consecutive measurements. Since the hydraulic state of the injector was uncertain when starting the FIE, the very first injection of fuel was never recorded. The injection duration was kept constant at 8 ms for all test points. This extended injection duration was used intentionally to facilitate the analysis of the steady state part of the injection separately from the transient phases linked with the needle opening and closure. Typical injection durations for this injector model can last up to 2 ms at medium and high engine load condition [28].

The data acquisition was performed using the manufacturer's software package, but to allow full control over the data analysis the raw measurements were exported from the LDV and processed using bespoke MATLAB scripts. The data acquisition system used in this study was able to record 1024 velocity and pressure samples over a time period of 10 ms. This represents a sampling time of 9.766 $\mu$s, or sampling frequency of 102 400 Hz, which meant that the highest frequency that could be resolved by discrete Fourier transform was 51.2 kHz, with a resolution of 100 Hz. In order to observe the evolution of the vibration spectrum throughout the injection event, spectrograms were produced using a sliding discrete short-time Fourier transform. The sliding FFT was applied on blocks of 64 data points with an overlap of 63 samples and a Hanning window, giving an effective resolution of 800 Hz for the spectrograms. The continuous spectrograms permit the differentiation of the frequencies that are specific to the needle opening and closing phases, from those that are observed throughout the injection event. The five individual spectrograms were then averaged, and the final step in the data processing was to eliminate low intensity noise by a applying a lower threshold to the averaged spectrograms' intensity. We found that the nozzle vibration frequencies in the X, Y and Z directions were sufficiently similar that the analysis could be performed in the Z direction only (i.e. along the injector axis). Hence all spectra are reported along that direction, and were computed based on the combination of the measurements by the three individual LDV heads sensors to produce the Z component. A conclusion from this work is that our approach could be simplified to a 1D LDV system in order to reduce the experimental setup's complexity, or in facilities which have limited optical access.

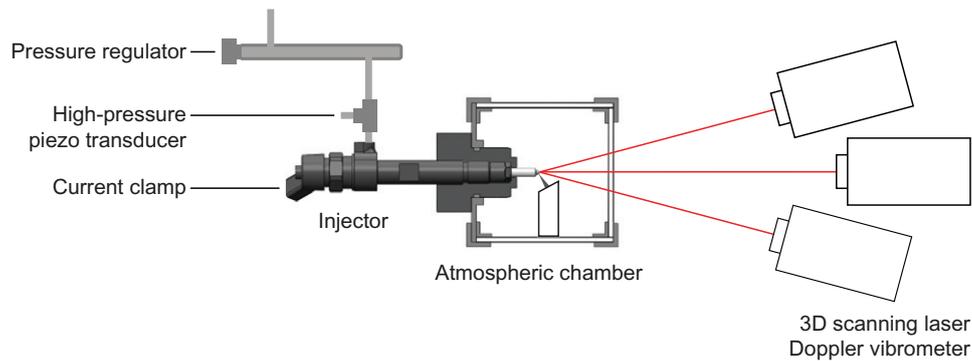

**Figure 4.** Schematic of the experimental configuration for the time-resolved 3D laser Doppler vibrometry on the atmospheric chamber.

*2.3. Vibration interferences and measurement uncertainties*

Although care was taken to minimise potential sources of interference, a number of cyclic and random phenomena could still degrade or interfere with the measurement of the spectral density of the nozzle tip. Since the frequency and amplitude of the nozzle vibration were obtained by measuring the Doppler shift of three laser beams focused on the nozzle tip, any spurious back scattering or beam steering could deteriorate the measurements. The presence of atomised fuel inside the static enclosure was indeed found to significantly interfere with the LDV measurements. This was established to be due to random light scattering by diesel droplets, and refraction of the laser beams by fuel deposited on the front window of the static enclosure. This issue was eliminated by capturing the spray with a 20 mm cylindrical pipe positioned downstream of the nozzle, and regularly cleaning the static enclosure's front window. We found that the noise generated by a disruption of the optical path was random rather than quasi-cyclic. The result is that the computed spectrograms contain broadband noise with no discernible structure, and thus they can easily be identified and rejected. An advantageous consequence of the LDV technique's high sensitivity to changes in refractive index is that it provides a strong and unambiguous indicator of an acquisition's contamination.

The common-rail system included several rotating and vibrating components that were expected to contribute to the power spectrum recorded by the LDV. A characterisation of these components was required so that the injector's contribution to the spectral density could be obtained. The first rotating component in the injection system was the low pressure pre-supply pump which drew fuel from the tank. The high-pressure pump's electric motor and three-plunger design introduced contributions in the spectral density at 47 Hz and 141 Hz, respectively. It is worth mentioning that some Bosch high-pressure pump models (e.g. CP1H and CP3) use an inlet metering valve (IMV), which regulates the quantity of compressed fuel according to the system demand [28]. The IMV reduces the demand on the pump and reduces the fuel temperature. The IMV is controlled using a 10 kHz signal, which could potentially introduce high-frequency pressure waves in the system and interfere with the spectral analysis. In our case, the high-pressure pump was an earlier model variant (CP1) which did not include an IMV. The regulation of the fuel pressure was done using a pressure regulator valve, mounted on the common-rail, which required a 1 kHz control signal.

In order to reduce the transmission of vibration from the pre-supply and high-pressure pumps, the injection system and static enclosure were mounted on separate tables with damped legs. The only rigid connection between these two tables was the high-pressure fuel line that linked the common-rail to the injector. The vibrations transmitted from the fuel pumps and ancillary equipment to the injector were characterised by recording baseline measurements. These were conducted by recording the vibration spectrum of the injector nozzle with the test bed in normal operation, but with no injection trigger pulse. This was repeated five times for every injection pressure, and the measurements averaged to produce baseline spectrograms.

The vibrometer could resolve frequencies up to 80 kHz and velocities between $0.01\,\mu\mathrm{m\,s^{-1}}$ and $1\,\mathrm{m\,s^{-1}}$. The measurement uncertainty is a function of the vibrometer's velocity decoder card, its measurement range, and the measured frequency. In our experiments the digital velocity decoder was a Polytech VD-03 with a measurement range of $100\,\mathrm{mm\,s^{-1}\,V^{-1}}$. Based on the instrument's specifications, the attainable resolution was better than $3 \times 10^{-6}\,\mathrm{m\,s^{-1}\,Hz^{0.5}}$, and typically of the order of $0.5 \times 10^{-6}\,\mathrm{m\,s^{-1}\,Hz^{0.5}}$.

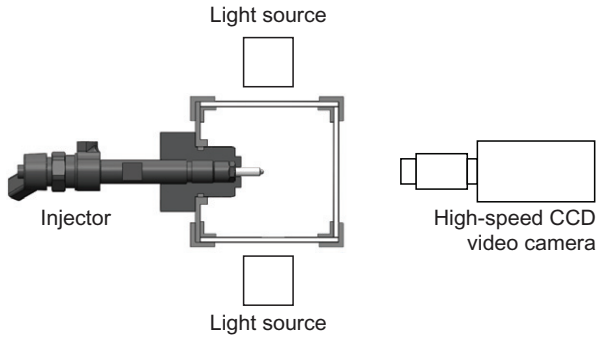

**Figure 5.** Schematic of the atmospheric chamber and high-speed video camera.

As mentioned previously low intensity noise was eliminated by a applying a lower threshold to the averaged spectrograms' intensity. The value of this lower threshold was arbitrarily set to $0.01\,\mathrm{m\,s^{-1}\,Hz^{0.5}}$.

### 2.4. High-speed video of the near-nozzle liquid jets

High-speed videos of the near-nozzle region were recorded for a range of conditions to measure the frequency of the liquid fuel sprays' transversal oscillations. The high-speed camera used for this series of experiments was a Kodak Ektapro HS Motion Analyzer (model 4540). The sprays were side-lit using a halogen flood light fitted with a diffuser (figure 5). The best compromise between acquisition rate and image resolution was obtained with a frame rate of 27 000 images per second, with a resolution of 0.412 mm per pixel. The initial 20 mm of the liquid jet were recorded for the VCO nozzle with a 200 μm orifice, with injection pressures of 60, 100 and 160 MPa. Amplitude spectra were produced from the videos by extracting the temporal evolution of the mean intensity of 3 × 3 pixel regions, and computing a fast Fourier transform. For each injection pressure, spectra were generated for five different locations at the periphery of the liquid jet.

## 3. Results and discussion

The spectrograms presented in figure 6 are averages of five individual measurements, taken for injection pressures of 60, 100, 120 and 160 MPa. In order to examine the nature and origins of the different acoustic phenomena, three different spectrograms are shown for each pressure condition: the nozzle tip without injection, the nozzle tip during injection, the pressure transducer during injection. LDV spectrograms for the nozzle tip without injection are presented in the left column of figure 6 to characterise the background parasitic noise generated by the ancillaries. They were recorded with the test bed in normal operation, but with the injection trigger disabled to prevent the injection of fuel. These spectrograms show that no significant vibration was generated by the test bed, with mild parasitic noise only below 3 kHz. LDV spectrograms for the nozzle tip vibration during injection are presented in the central column of figure 6. Spectrograms for the signal measured by the high-pressure transducer directly upstream of the injector are presented in the right column of figure 6. Since the mechanical operation of the injector is known to lead to fuel pressure fluctuations [24], computing the pressure transducer spectrograms can give an indication of the frequencies induced by the mechanical operation of the injector. Additionally, using the pressure sensor simultaneously to the LDV also allows triangulating the origin of some of the frequencies in the spectrograms, thus permitting an estimate of the source of vibration within the system.

### 3.1. Nozzle opening and closing

The nozzle tip spectrograms in figure 6 indicate that significant levels of low and high frequency vibrations occur during the first 0.5 ms of the injection period, for all injection pressures. Possible causes of vibration that may occur in diesel injectors during this transient phase include: the opening of the control volume by the solenoid valve, quasi-cyclic vortex shedding in the nozzle sac [30], fluid hammer induced by the prompt opening of the orifice [31] and quasi-cyclic cavitation in the nozzle orifices [32]. Firstly, the quasi-cyclic shedding of vortices is not expected to contribute significantly to the vibration spectra in figure 6 due to the sac-less nozzle design and the reduced flow recirculation expected in a single hole nozzle. Fluid hammer may occur at the beginning of the injection process if the nozzle is opened sufficiently rapidly [33]. Following the initial pressure drop due to the opening of the nozzle, a pressure rise may then occur due to fluid hammer waves reflected by the edge filter located at the injector's inlet fitting (figure 1). The following criterion can be used to evaluate the shortest opening time that will cause fluid hammer to occur at the outlet of a duct, based on its length ($L$) and the speed of sound ($c$) [33]:

$$t = \frac{2L}{c} \qquad (1)$$

The full length of the injector's internal duct could not be measured directly, but was estimated as 177 mm based on external measurements and internal schematics. The values for the speed of sound were calculated using correlations in [34], for a diesel fuel with similar properties to ours, to obtain the nozzle opening time criteria ($t$) and fluid hammer frequencies ($f = 1/t$) for the pressure range in our experiment:

60 MPa   $c = 1450\,\mathrm{m\,s^{-1}}$   $t = 0.24\,\mathrm{ms}$   $f = 4.2\,\mathrm{kHz}$

160 MPa   $c = 1720\,\mathrm{m\,s^{-1}}$   $t = 0.21\,\mathrm{ms}$   $f = 4.8\,\mathrm{kHz}$

These times are smaller than the time required for the needle to reach full lift, suggesting that fluid hammer may not take place during nozzle opening. However, we cannot fully reject this eventuality as the complex geometry of the injector's fuel lines means that its spectral response will deviate from that of a perfectly cylindrical duct. We also note that the calculated fluid hammer frequencies correlate well with a peak consistently found in the bottom left corner of all nozzle tip spectrograms (time < 0.5 ms, frequency < 5 kHz).

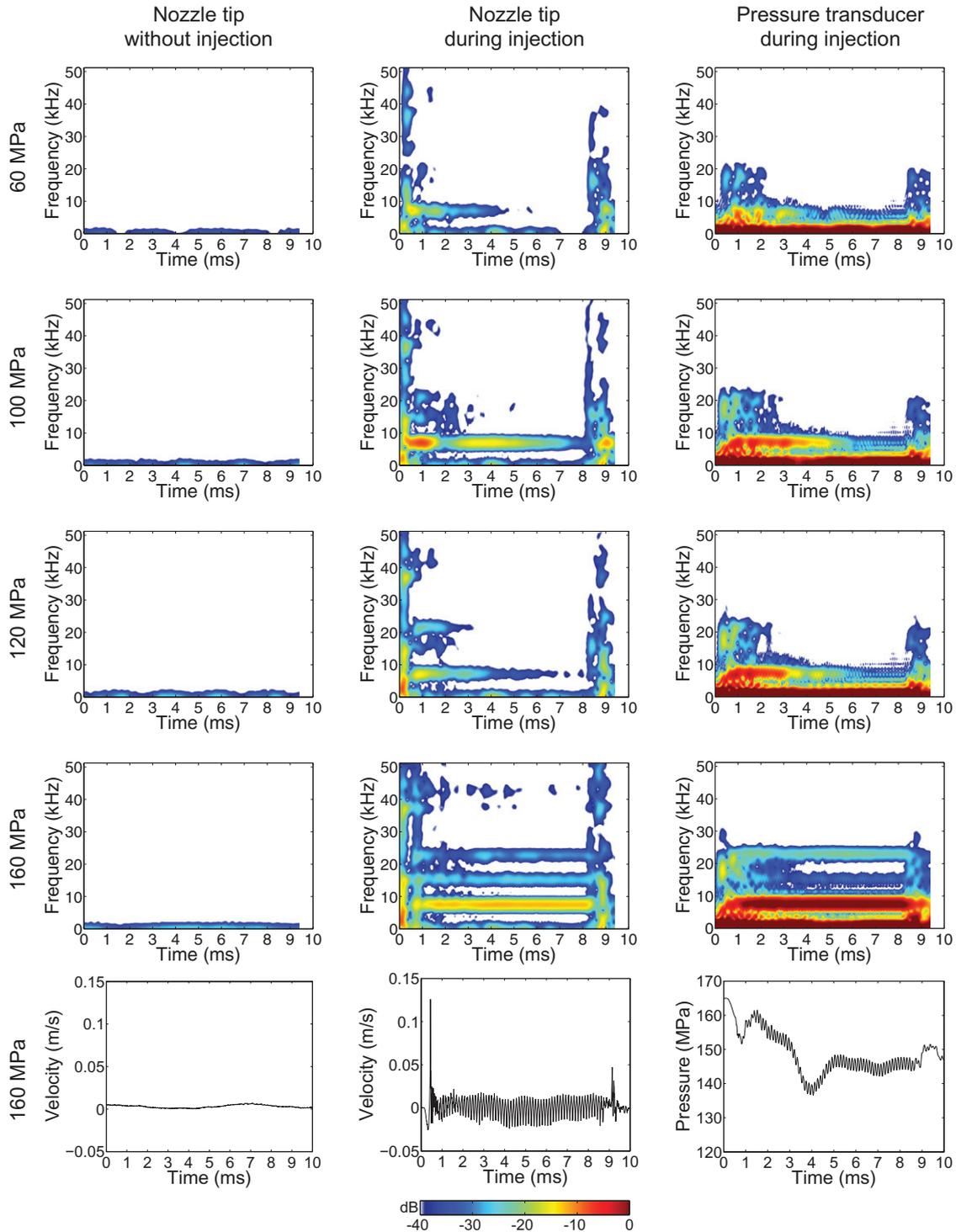

**Figure 6.** Averaged spectrograms for a VCO 200 $\mu$m single-hole nozzle, recorded by LDV without injection (left), by LDV during injection (centre) and by the pressure transducer upstream of the injector, during injection (right). Raw signals in the time domain are presented at the bottom of the table for 160 MPa. Intensity scales are in dB ref m s$^{-1}$ Hz$^{-0.5}$ for the LDV spectrograms, and dB ref Pa Hz$^{-0.5}$ for the pressure transducer spectrograms.

The opening of the control volume by the solenoid valve is likely to generate mechanical and hydraulic vibrations before the start of injection. This explains why oscillations can be observed for early timings in figure 6. Since the control volume is located closer to the fuel pressure sensor than to the nozzle tip, we can expect the vibrations generated by its opening to be visible in the pressure transducer spectrograms. Since these show no significant peaks above 20 kHz, we can conclude that the opening of the control volume generates vibrations below this value. There is only one peak that appears simultaneously on the LDV and pressure transducer spectrograms during the first 0.5 ms, with a frequency of 15 kHz, suggesting that this frequency might be associated with the solenoid valve opening the control volume.

The first 0.5 ms of the injection period time is also dominated by the transient processes that take place during the needle opening, which lead to increased levels of turbulence [35] and cavitation in the nozzle [36]. Since these vibrations will take place close to the LDV sampling point, we can expect them to be visible in the nozzle tip spectrograms. Considering that acoustic attenuation is an exponential function of both the path length and the square of the frequency, high frequency hydraulic vibrations originating inside the nozzle are unlikely to be measured by the pressure transducer located upstream of the injector. Hence by comparing the LDV and pressure spectrograms in figure 6 we can conclude that vibrations with frequencies higher than 25 kHz are related to quasi-cyclic turbulence phenomena taking place inside the nozzle, which may include cavitation. This point will be further elaborated in the next section. From the previous discussions, vibrations that occur during nozzle opening with frequencies lower than 25 kHz are related to the opening of the control volume by the solenoid valve (15 kHz) and possibly fluid hammer induced by the rapid opening of the orifices (3–5 kHz). It is possible that cavitation and turbulence also generate frequencies lower than 25 kHz, although there was no indication that this was the case in our experiment.

The augmented level of vibration was particularly noticeable during the closing phase (between 8 and 8.5 ms in figure 6). We obtained evidence that these vibrations occurred at the nozzle by triangulating the recordings from the LDV and the fuel pressure sensor. In the case of 60 MPa injection pressure (top row in figure 6), broadband vibrations were first recorded by the LDV at 0.04 ms for the nozzle opening, and at 8.12 ms for the nozzle closing event. The same broadband vibrations reached the pressure sensor at 0.25 ms and 8.35 ms, respectively. These timings confirm that the origin of the vibrations is closer to the LDV than the pressure sensor, and therefore that they must originate nearer the tip of the injector than its fuel inlet. Knowing the speed of sound of diesel fuel at 60 MPa (1450 m s$^{-1}$) and the distance between the LDV and pressure sensor, one can use the time at which the two instruments register the same vibration events to estimate the location at which the vibration initiated. The time difference between these two instruments measuring the same events indicated that the source of vibration was within the nozzle tip. This is consistent with findings that with low needle lifts the turbulent kinetic energy is increased [35, 37] and cavitation structures are more unstable [35]. Salvador *et al* [37]. suggested that the flow area contraction during needle descent causes an acceleration of the fuel, thus generating a strong pressure drop and phase change. An alternative explanation by Lee and Reitz [38] is that fuel pressure drops during the nozzle closing phase, as the fuel flow to the nozzle orifices becomes restricted by the needle but the exit mass flow decreased relatively slowly due to inertia, leading to increased cavitation inside the orifices. Significant levels of low and high frequency vibrations occurred at the time of nozzle closure, between 8.5 and 9 ms, for all injection pressures. These can be attributed to fluid hammer, which is known to be generated during injector closure due to the sudden change in fuel flow velocity [39, 40].

### 3.2. Steady-state injection phase

For the purpose of this discussion, the steady-state injection phase refers to times between 0.5 and 8 ms after start of injection trigger, when the rate of injection is approximately constant with time. We observe that LDV spectrogram in figure 6 for 160 MPa fuel pressure shows intense vibrations between 0.7 and 1.0 ms. This timing coincides with the needle reaching its maximum lift (figure 2), suggesting that these vibrations may be linked to the needle plunger hitting its mechanical stop. In the case of servo-actuated injectors, such as the one used in this study, the displacement of the needle is determined by the pressure difference between the fuel around the needle and the fuel in the servo valve control volume. For non-ballistic injectors and if the injection duration is sufficiently long, as is the case for medium and high engine load, the injector's needle plunger reaches its full lift and dwells on the film of fuel that flows through the valve control volume [28]. The rapid lifting of the needle results in high kinetic energy which must be dissipated through mechanical deformation of the plunger and needle, or transferred to the fuel. Hence for high injection pressures the plunger cannot settle instantaneously at hydraulic stop but rebound from mechanical stop, thus generating intense pressure waves [41]. The vibrations generated by this impact appear between 0.7 and 1.0 ms in some of the LDV spectrograms.

The nozzle tip LDV spectrograms in the central column of figure 6 all indicate an intense oscillation with a frequency of approximately 7 kHz, which is initiated between 0.7 and 1.0 ms and extends into the steady part of the injection. For the lowest injection pressure this vibration is damped and disappears before 5 ms. The dominant frequency in the pressure spectrograms is also 7 kHz, in agreement with the LDV measurements. For the highest injection pressure this frequency remains until the end of injection and, notably, shows no sign of damping, neither in the LDV nor the pressure-based spectrograms. Similarly, a spectral analysis of needle plunger lift measurements showed that the main frequency at which the injector needle oscillates in the axial direction is close to 7 kHz. The exact value of this frequency peak was found to be linearly correlated to the injection pressure (figure 7). The frequency peaks were obtained from power spectral density plots, such as the one shown in figure 8, with a resolution of 100 Hz for the LDV and pressure data. With this resolution the trend of vibration frequency with rail pressure is mild but significant. It can be observed that the mild sensitivity of the fuel pressure frequency to injection pressure is also consistent with the correlations observed for the LDV. This frequency was also found to be proportional to the injection pressure, and measured between 5.5 and 7.2 kHz for pressures between 60 and 160 MPa. Interestingly, the second and third harmonics of the 7 kHz fundamental frequency are present on some of the LDV and pressure spectrograms in figure 6. These harmonics are particularly visible at 160 MPa with bands visible at 7.7, 15 and 23 kHz throughout the injection time. The presence of harmonics indicates that part of the hydraulic system is in resonance. As the harmonics are more intense on the LDV than on the pressure-based spectrograms, the injector must be part of the resonating system. The 3D LDV data showed that this intense 7 kHz oscillation frequency was recorded along all three axes, which is consistent with a resonance phenomenon. In order to establish if other components such as the high-pressure fuel lines or common rail are part of the resonating system, we can use the following formula to calculate the length ($L$) of the system based on its natural frequency ($f$):

$$L = \frac{c}{2f} \qquad (2)$$

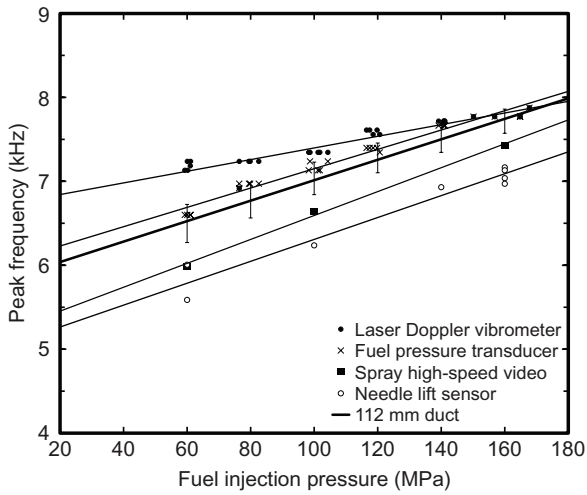

**Figure 7.** Correlations between injection pressure and 7 kHz oscillation peak frequency shift measured for the injector tip (LDV), fuel pressure (piezoresistive transducer), near nozzle liquid fuel spray (high-speed video) and injector needle (plunger lift sensor). The natural frequency of a 112 mm open-ended duct is included for comparison, with error bars indicating the effect of a ±20 K change in fuel temperature.

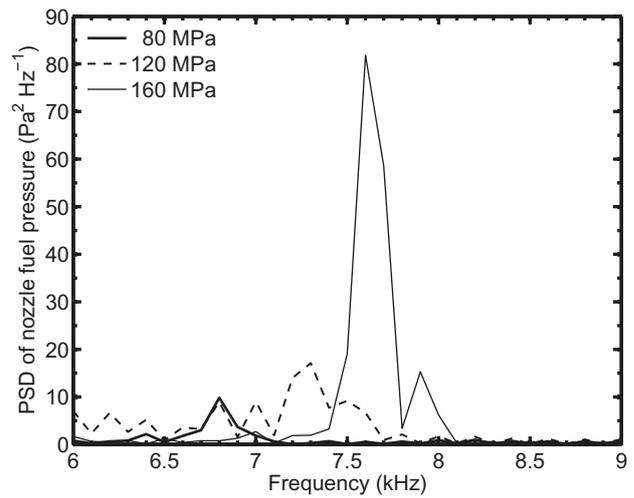

**Figure 8.** Power spectral density plots of the pressure transducer signal, for different fuel injection pressures. The amplitude and frequency of the main peaks can be seen to increase with pressure. Each line plot is the average of five consecutive measurements.

By calculating the speed of sound ($c$) using the correlations in [34] for diesel fuel at 160 MPa pressure, and using $f = 7.7$ kHz for the natural frequency, we can estimate the length of the resonating system as 110 mm. We can observe that this value corresponds almost exactly to the length of the cylindrical fuel line that links the injector inlet to the needle (figure 1). In order to substantiate the theory that this fuel line is in resonance, figure 7 shows the natural frequency of a 112 mm open-ended duct as a function of fuel pressure (and therefore speed of sound), compared to the evolution of the 7 kHz peak measured by the other instruments. We can observe that the frequencies measured by the pressure sensor are in agreement with the natural frequency of a 112 mm duct. The agreement with the LDV measurements is good for the higher fuel pressures but deviates by up to 10% at the lower pressures, indicating that other mechanisms than resonance are also affecting the injector's response. This appears consistent with the fact that the vibration at the natural frequency of the injector's main fuel line is damped for the lower fuel pressures and undamped for 160 MPa. It is worth stressing that the length of the internal fuel line will vary between injector models, and therefore the natural frequency will change accordingly. Other second generation Bosch common rail injector can have an elongated geometry, with an internal fuel line longer than the one used in our experiments [42]. For example Payri and Manin [31, 43] used a similar injector but with an overall length of 130 mm for the internal fuel line, and reported oscillation frequencies between 6 and 7 kHz from rate of injection, spray momentum and accelerometer measurements. They suggested this was due to the needle-rod-fluid system reaching resonance. Applying our reasoning to their experiment, the natural frequency of a 130 mm open-ended line filled with n-dodecane at 343 K and 150 MPa is 6.8 kHz, in agreement with the frequencies reported in [43] and present in figure 7 of [31] between 1.5 and 3 ms.

Whilst our measurements showed a correlation between internal flow oscillations and external nozzle tip vibration, the transmission of these cyclic fluctuations to the liquid jets remained to be ascertained. High-speed videos of the fuel sprays were recorded to investigate whether nozzle tip vibrations could be transmitted to the near nozzle liquid jet. Frames extracted from a high-speed video of a diesel spray are shown in figure 9 to illustrate the cyclic oscillations in the appearance of the near-nozzle jet periphery. Initial estimates of the frequency of these oscillations were obtained by calculating the time difference between two frames showing similar spray shapes [26]. The frequency was found to be 6.8 kHz for a single-hole VCO nozzle with a 200 μm orifice diameter and 160 MPa fuel pressure. In the present study, the mean intensity of 3 × 3 pixels regions near the spray interface were measured for each video frame, between 5 and 20 mm from the orifice. These temporal intensity fluctuations were then used to calculate the spectral density for each region, using a fast Fourier transform. It can be observed in figure 7 that the frequencies measured for the liquid jet oscillations are reasonably close to those measured with the other instruments. Notably, the linear correlation between fuel pressure and oscillation frequency was also observed in the high-speed video spectra, thus supporting the fact that nozzle vibrations are transmitted to the liquid jet. The jet oscillation frequencies are between 0.5 and 1 kHz lower than the LDV measurements. This discrepancy could be partially explained by differences in fuel temperature (and therefore speed of sound) between the LDV and high-speed video experiments. The effect of a ±20 K difference in fuel temperature on the natural frequency of an open-ended duct are ±230 Hz at 60 MPa and ±140 Hz at 160 MPa. Beyond the effect of temperature, a level of damping between the injector's vibration and the oscillation of the liquid fuel spray is to be expected. Therefore the vibrations measured by vibrometry should be seen as an upper limit for the spray oscillations measured by video. Vortex shedding at the tip of fuel injector needles is known to occur at high frequencies [30, 44] and can lead to cyclic fluctuations in the shape of near nozzle fuel sprays [45]. Our results show that pressure waves generated by resonance of the injector fuel line may also lead to cyclic fluctuations in the near nozzle fuel sprays, in agreement with [31]. CFD predictions have indicated that such transversal spray oscillations lead to temporal and spatial clustering of the droplets in the fuel plume [27]. These inhomogeneous liquid fuel distributions should inhibit soot oxidation due to the associated local depletion of oxygen and lead to clustered distributions of soot, in agreement with experimental observations [46].

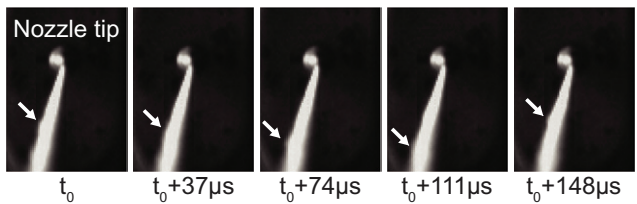

**Figure 9.** Close-up images obtained by high-speed video for a VCO 200 μm single-hole nozzle, showing the oscillation of the liquid spray. Similar spray patterns were observed every 148 μs and are indicated by the white arrows. Frame field of views are 26 × 37 mm, injection pressure was 160 MPa, in-cylinder pressure was 5.2 MPa.

Vibrations are also expected to occur during the injection process due to cavitation taking place inside nozzle orifices, as well as unsteady turbulent flow phenomena such as vortex shedding at the needle tip and nozzle sac [30, 44]. The growth and collapse of cavitation should lead to fluctuations in fuel pressure and mass flow rate inside the injector and nozzle orifices. The spectra for internal nozzle flow could not be measured independently from the rest of the system, but it is proposed that they can be inferred by subtracting the pressure transducer spectrograms from the total nozzle vibration spectrograms shown in the right and central columns of figure 6, respectively. This is justified by the fact that acoustic attenuation is an exponential function of the path length, hence vibrations originating inside the nozzle are unlikely to be measured by the fuel pressure transducer. It is well established that higher injection pressures, or more generally higher Reynolds numbers, increase the tendency for cavitation to be more pronounced [47]. The spectrograms in the central column of figure 6 show signs of increased vibrations as the pressure is raised from 60 to 160 MPa. Indeed, vibrations occurring between 35 and 45 kHz extend further into the steady state regime of the injection as pressure increases. This is particularly noticeable at 160 MPa, where a 45 kHz oscillation is visible throughout the injection period. Whilst there is a lack of experimental data for the frequency of cyclic growth and collapse of cavitation bubbles in realistic high pressure diesel injectors, numerical simulations by Schmidt [32] predict such frequencies to be in the order of 26 to 33 kHz for a VCO nozzle and 100 MPa fuel pressure. Other numerical simulations suggest that cyclic cavitation processes may occur at even higher frequencies, beyond the measurement range of our experiments. Notably, simulations by Payri [48] indicate that the growth and collapse of vapour cavities may occur cyclically with a period of 11 μs, equivalent to a frequency of 90.9 kHz, with an injection pressure of 80 MPa and back pressure of 3 MPa. Even in the case of stable cavitation, when the length of the cavity is almost constant with time, vibrations may be generated as the wake region always fluctuates [49]. Simulations by Yuan [50] indeed suggest that high frequency oscillations may be found when the wake region of the cavitation oscillates, with the vapour fraction in the nozzle changing at a frequency in excess of 120 kHz.

### 3.3. Effect of nozzle geometry

The effect of nozzle geometry on the vibrational properties of the injector was characterised for a range of orifice diameter and nozzle design. Figure 10 shows the spectrograms recorded for VCO, minisac and microsac nozzles, with orifices diameters of 100 and 200 μm. The spectrograms do not show major differences during the nozzle opening phase, between 0 and 0.5 ms. All nozzle geometries generated similarly high levels of vibration, with a spectrum of frequencies that extended beyond 50 kHz. The similarity between the spectral responses for these nozzles during these early timings supports the previous discussions that vibrations taking place during the opening phase are mainly generated by fluid hammer and the operation of the solenoid valve, both effects being expected to be insensitive to nozzle geometry.

Since all orifices were cylindrical we expect that all nozzles generated cavitating flows. However, the change in nozzle geometry and orifice diameter should affect the mass flow and Reynolds number, and therefore the degree of cavitation [47] both at the needle tip and inside the orifices. Thus, considering both the L/D ratios and orifice diameters for the nozzles we tested (table 1), we can reasonably expect that the orifices with the largest L/D ratio should cavitate less. Intermittent vibrations with frequencies between 35 and 45 kHz were visible in figure 6 for a VCO nozzle with an injection pressure of 160 MPa. Similar sporadic vibrations can also be observed in figure 10 for the minisac and microsac nozzles throughout the injection period, although only for the 200 μm orifices. The 100 μm orifices show little evidence of vibrations within that frequency range, supporting the hypothesis that the 35–45 kHz vibrations seen on the LDV spectrograms for larger orifices are an indication of cavitation inside the nozzles. Compared to the VCO nozzles, the minisac and microsac nozzles lead to more intense, high frequency, vibrations from 8 ms after start of injection trigger. This may be caused by a pressure drop inside the nozzle sac and orifices during the needle closure, leading to more pronounced cavitation. Lee and Reitz [38] conducted simulations of transient cavitation processes for sac-type diesel injectors, which predicted increased levels of cavitation during the closing phases. They suggested that cavitation increases due to rarefaction, caused by transient differences between the mass flow rate of fuel entering the nozzle sac and the mass flow rate of fuel exiting the nozzle orifice. This was explained by the needle rapidly restricting the inlet flow while inertia at the orifice outlet results in a higher exit mass flow rate [38], thus leading to a pressure drop inside the nozzle orifices and increased levels of cavitation.

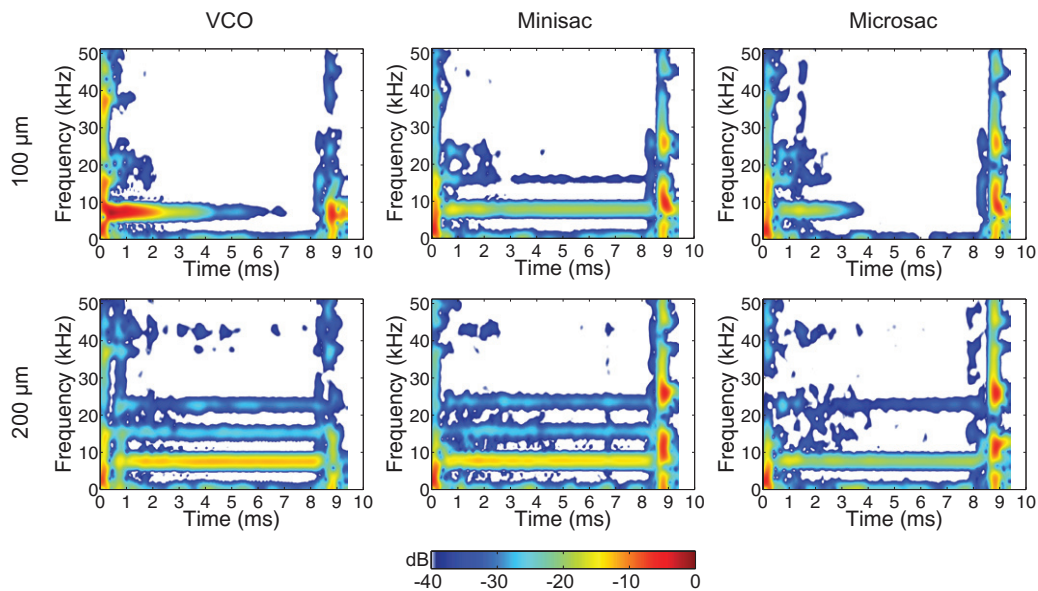

**Figure 10.** Spectrograms for single-hole VCO, minisac and microsac nozzles, with diameters of 100 μm (top row) and 200 μm (bottom row) recorded for injections with 160 MPa pressure. Intensity scales are in dB ref m s$^{-1}$ Hz$^{-0.5}$.

## 4. Conclusions

We investigated the internal flow of diesel fuel injectors by simultaneously measuring the nozzle tip vibration, upstream fuel pressure and needle lift. The technique we propose, based on the triangulation of a laser Doppler vibrometer and fuel pressure transducer signals, enables the quantitative characterisation of internal flows for unmodified injectors with real fuels at elevated injection pressures. The simultaneous spectral analysis of 3D LDV and fuel pressure sensor data can provide quantitative time-resolved information for quasi-cyclic mechanical and hydraulic phenomena inside fuel injectors. The technique enables indirect and non-intrusive investigations of internal cyclic processes, such as cavitation, vortex shedding and turbulent nozzle flow, inside real-size injectors.

It can also be used with real fuels at elevated injection pressures, and does not require physical modifications of the nozzles. Hence, there is a wide scope for future works both on the development of the technique and its application to multi-hole fuel injectors at elevated in-cylinder pressures and temperatures. Our results suggest that the approach we propose could be reduced to a 1D LDV head if optical access is limited, or to reduce the experimental setup's complexity. The measurement range of the laser Doppler vibrometer could be extended beyond 51 kHz to investigate the occurrence of faster processes.

We applied our approach to compute time-resolved spectrograms of a diesel injectors' vibrations during injection events, for a range of nozzle geometries and injection pressures. An oscillation between 6 and 7.5 kHz was measured using the laser vibrometer and was found to be in agreement with results obtained from the fuel pressure transducer, injector needle lift sensor, and high-speed videos of the near-nozzle diesel jet. This frequency was recorded for all nozzles tested and was found to be proportional to the injection pressure by the four independent instruments. This frequency is proposed to be the natural frequency for the injector's main internal fuel line. Once resonance is initiated the needle and plunger start oscillating, further affecting the flow of fuel in the nozzle. A Fourier analysis of high speed videos recorded near the nozzle exit showed that these oscillations were transmitted to the liquid jet, potentially contributing to the formation of fuel droplet clusters.

By comparing the LDV and pressure spectrograms we conclude that vibrations with frequencies higher than 25 kHz are related to quasi-cyclic turbulence phenomena taking place inside the nozzle, which may include cavitation. Vibrations that occur during nozzle opening with frequencies lower than 25 kHz are related to the opening of the control volume by the solenoid valve (15 kHz) and fluid hammer induced by the rapid opening of the orifices (3 to 5 kHz depending on fuel pressure).

Oscillations between 35 and 45 kHz were observed during the needle opening phase for all nozzles and injection pressures. For the 200 $\mu$m nozzle orifices, these oscillations were found to progressively extend into the steady state injection period as the injection pressure was increased. This frequency band may indicate the presence of quasi cyclic cavitation inside the orifices.

Compared to the VCO nozzles, minisac and microsac nozzle types exhibited more intense vibrations during the needle closing phase, suggesting higher levels of cavitation. Vibrometric measurements could be coupled with high-speed videos of the internal nozzle flow to provide a direct link between nozzle vibration spectra and cyclic cavitation phenomena that occur inside the orifice and at the needle tip.

## Acknowledgments

This work was supported by the UK's Engineering and Physical Science Research Council [grant number EP/F069855/1]; and the European Regional Development Fund [INTERREG IVA grant number 4005]. The authors are grateful to Ricardo UK for supplying equipment, Thibault Panhard for his assistance with the experiment, and to the EPSRC Engineering Instrument Pool for the loan of equipment used for this study. The authors wish to thank the four anonymous reviewers for their constructive criticisms of an earlier version of the manuscript.

## References

[1] Sazhin S, Crua C, Kennaird D and Heikal M 2003 The initial stage of fuel spray penetration *Fuel* **82** 875–85

[2] Som S and Aggarwal S K 2010 Effects of primary breakup modeling on spray and combustion characteristics of compression ignition engines *Combust. Flame* **157** 1179–93

[3] Dumouchel C 2008 On the experimental investigation on primary atomization of liquid streams *Exp. Fluids* **45** 371–422

[4] Faeth G M, Hsiang L P and Wu P K 1995 Structure and breakup properties of sprays *Int. J. Multiphase Flow* **21** 99–127

[5] Gorokhovski M and Herrmann M 2008 Modeling primary atomization *Ann. Rev. Fluid Mech.* **40** 343–66

[6] Heimgärtner C and Leipertz A 2000 Investigation of primary diesel spray breakup close to the nozzle of a common rail high pressure injection system *8th Int. Conf. on Liquid Atomization and Spray Systems (ICLASS)* (Pasadena, USA)

[7] Hossainpour S and Binesh A 2009 Investigation of fuel spray atomization in a DI heavy-duty diesel engine and comparison of various spray breakup models *Fuel* **88** 799–805

[8] Lefebvre A 1989 *Atomization and Sprays* (New York: Hemisphere Publishing)

[9] Reitz R and Diwakar R 1987 Structure of high-pressure fuel sprays *SAE Technical Paper* 870598

[10] Duke D, Swantek A, Tilocco Z, Kastengren A, Fezzaa K, Neroorkar K, Moulai M, Powell C and Schmidt D 2014 X-ray imaging of cavitation in diesel injectors *SAE Int. J. Engines* **7** 1003–16

[11] Duke D J, Kastengren A L, Tilocco F Z, Swantek A B and Powell C F 2013 X-ray radiography measurements of cavitating nozzle flow *Atomization Sprays* **23** 841–60

[12] Hayashi T, Suzuki M and Ikemoto M 2013 Effects of internal flow in a diesel nozzle on spray combustion *Int. J. Eng. Res.* **14** 646–54

[13] Arcoumanis C, Gavaises M, Flora H and Roth H 2001 Visualisation of cavitation in diesel engine injectors *Mec. Ind.* **2** 375–81

[14] Gavaises M, Andriotis A, Papoulias D, Mitroglou N and Theodorakakos A 2009 Characterization of string cavitation in large-scale Diesel nozzles with tapered holes *Phys. Fluids* **21** 052107

[15] Sato K and Saito Y 2001 Unstable cavitation behaviour in circular-cylindrical orifice flow *CAV 2001: Fourth Int. Symp. on Cavitation (Pasadena, USA)* (http://resolver.caltech.edu/CAV2001:sessionA9.003)

[16] Schmidt D P, Rutland C J, Corradini M, Roosen P and Genge O 1999 Cavitaton in 2D asymmetric nozzles *SAE Technical Paper* 1999-01-0518

[17] Roosen P, Unruh O and Behman M 1997 Investigation of cavitation phenomena inside fuel injector nozzles *Int. Symp. on Automotive Technology and Automation (ISATA) (Florence, Italy)*

[18] Laoonual Y, Yule A J and Walmsley S J 2001 Internal fluid flow and spray visualization for a large scale valve covered orifice (VCO) injector nozzle *17th Eur. Conf. on Liquid Atomization and Spray Systems (ILASS) (Zurich, Switzerland)*

[19] Stutz B and Reboud J L 1997 Experiments on unsteady cavitation *Exp. Fluids* **22** 191–8

[20] Sato K and Shimojo S 2003 Detailed observations on a starting mechanism for shedding of cavitation cloud *CAV 2003: Fifth Int. Symp. on Cavitation (Osaka, Japan)* (http://iridium.me.es.osaka-u.ac.jp/cav2003)

[21] Martynov S 2005 Numerical simulation of the cavitation process in diesel fuel injectors *PhD Thesis* University of Brighton, UK

[22] Franc J P 2001 Partial cavity instabilities and re-entrant jet *CAV 2001: 4th Int. Symp. on Cavitation (Pasadena, USA)* (http://resolver.caltech.edu/CAV2001:lecture.002)

[23] Wang X and Su W 2010 Numerical investigation on relationship between injection pressure fluctuations and unsteady cavitation processes inside high-pressure diesel nozzle holes *Fuel* **89** 2252–9

[24] Chaves H and Obermeier F 1996 Modelling the effect of modulations of the injection velocity on the structure of diesel sprays *SAE Technical Paper* 961126

[25] Pontoppidan M, Ausiello F, Bella G and Ubertini S 2004 Study of the impact on the spray shape stability and the combustion process of supply pressure fluctuations in CR-Diesel injectors *SAE Technical Paper* 2004-01-0023

[26] Crua C 2002 Combustion processes in a diesel engine *PhD Thesis* University of Brighton, UK

[27] Katoshevski D, Shakked T, Sazhin S S, Crua C and Heikal M R 2008 Grouping and trapping of evaporating droplets in an oscillating gas flow *Int. J. Heat Fluid Flow* **29** 415–26

[28] Dietsche K-H, Klingebiel M and Müller R (eds) 2005 *Diesel Fuel-Injection System Common Rail* 2nd edn (Plochingen: Robert Bosch GmbH)

[29] Castellini P, Martarelli M and Tomasini E P 2006 Laser Doppler vibrometry: development of advanced solutions answering to technology's needs *Mech. Sys. Signal Proc.* **20** 1265–85

[30] Andriotis A, Gavaises M and Arcoumanis C 2008 Vortex flow and cavitation in diesel injector nozzles *J. Fluid. Mech.* **610** 195–215

[31] Manin J, Kastengren A and Payri R 2012 Understanding the acoustic oscillations observed in the injection rate of a common-rail direct injection diesel injector *J. Eng. Gas Turb. Power* **134** 122801

[32] Schmidt D P 1997 Cavitation in diesel fuel injector nozzles *PhD Thesis* University of Wisconsin, USA
[33] Parmakian J 1963 *Waterhammer Analysis* (New York: Dover Publications)
[34] Payri R, Salvador F J, Gimeno J and Bracho G 2011 The effect of temperature and pressure on thermodynamic properties of diesel and biodiesel fuels *Fuel* **90** 1172–80
[35] Roth H, Gavaises M and Arcoumanis C 2002 Cavitation initiation, its development and link with flow turbulence in diesel injector nozzles *SAE Technical Paper* 2002-01-0214
[36] Margot X, García A, Fajardo P and Patouna S 2010 Analysis of the cavitating flow in real size diesel injectors with fixed and moving needle lift simulations *5th Eur. Conf. on Computational Fluid Dyanmics (ECCOMAS CFD 2010) (Lisbon, Portugal)*
[37] Salvador F J, Martínez-López J, Caballer M and De Alfonso C 2013 Study of the influence of the needle lift on the internal flow and cavitation phenomenon in diesel injector nozzles by CFD using RANS methods *Energ. Convers. Manage.* **66** 246–56
[38] Lee W G and Reitz R 2009 Simulation of transient cavitation processes in diesel injectors using KIVA with a Homogeneous Equilibrium Model *Int. Multidimensional Engine Modeling User's Group Meeting (Detroit, USA)*
[39] Seykens X L J, Somers L M T and Baert R S G 2004 Modelling of common rail fuel injection system and influence of fluid properties on injection process *Int. Conf. on Vehicles Alternative Fuel Systems and Environmental Protection (Dublin, Ireland)*
[40] Hu Q, Wu S F, Stottler S and Raghupathi R 2001 Modelling of dynamic responses of an automotive fuel rail system, part 1: injector *J. Sound Vib.* **245** 801–14
[41] Spurk J, Betzel T and Simon N 1992 Interaction of nonlinear dynamics and unsteady flow in fuel injectors *SAE Technical Paper* 920621
[42] Payri R, Salvador F J, Martí-Aldaraví P and Martínez-López J 2012 Using 1D modeling to analyse the influence of the use of biodiesels on the dynamic behavior of solenoid-operated injectors in common rail systems: Detailed injection system model *Energ. Convers. Manage.* **54** 90–9
[43] Payri R and Manin J 2011 Hydraulic characterization *Engine Combustion Network Workshop 1 (ECN1)*
[44] Shi J-M, Wenzlawski K, Helie J, Nuglisch H and Cousin J 2010 URANS and SAS analysis of flow dynamics in a GDI nozzle *23rd Eur. Conf. on Liquid Atomization and Spray Systems (ILASS) (Brno, Czech Republic)*
[45] Zigan L, Schmitz I, Wensing M and Leipertz A 2012 Reynolds number effects on atomization and cyclic spray fluctuations under gasoline direct injection conditions *Fuel Systems for IC Engines* (London, UK: Woodhead Publishing) pp 253–63
[46] Crua C, Kennaird D A and Heikal M R 2003 Laser-induced incandescence study of diesel soot formation in a rapid compression machine at elevated pressures *Combust. Flame* **135** 475–88
[47] Payri F, Bermúdez V, Payri R and Salvador F J 2004 The influence of cavitation on the internal flow and the spray characteristics in diesel injection nozzles *Fuel* **83** 419–31
[48] Payri F, Payri R, Salvador F J and Martínez-López J 2012 A contribution to the understanding of cavitation effects in Diesel injector nozzles through a combined experimental and computational investigation *Comput. Fluids* **58** 88–101
[49] Coutier-Delgosha O, Reboud J L and Delannoy Y 2003 Numerical simulation of the unsteady behaviour of cavitating flows *Int. J. Numer. Methods Fluids* **42** 527–48
[50] Yuan W, Sauer J and Schnerr G H 2001 Modeling and computation of unsteady cavitation flows in injection nozzles *Mec. Ind.* **2** 383–94